\documentstyle[aps,epsf,graphicx]{revtex}
\def \D {\mbox{D}}
\begin{document}

\twocolumn[\hsize\textwidth\columnwidth\hsize\csname
@twocolumnfalse\endcsname

\title{Singularities on the brane aren't isotropic}
\author{
Marco Bruni$^1$  and Peter K. S. Dunsby$^{2,3}$}
\address{$1$ Institute of Cosmology and Gravitation,
University of Portsmouth, Mercantile House, Portsmouth PO1 2EG, U.K.\\
$2$ Department of Mathematics and Applied Mathematics,
University of Cape Town, Rondebosh 7701, Cape Town, South Africa.\\
$3$ South African Astronomical Observatory, Observatory
7925, Cape Town, South Africa.}
\date{July 20, 2002}

\newcommand{\beq}{\begin{equation}}
\newcommand{\eeq}{\end{equation}}
\newcommand{\lab}{\label}
\newcommand{\dd}{{\rm d}}
\newcommand{\ee}{{\rm e}}
\newcommand{\tnabla}{\tilde{\nabla}}
\newcommand{\hs}{\,-\,}

\maketitle

\begin{abstract}
Recent studies of homogeneous anisotropic universe models in the brane
world scenario show that the cosmological singularity in this context
is isotropic. It has therefore been suggested that this may be a generic 
feature of singularities on the brane, even in the inhomogeneous case. 
Using a perturbative approach, we show that this is not the case.
As in the GR case, the presence of decaying modes in the perturbations
signal the instability (in the past) of the isotropic singularity.
The brane universe is therefore not born with isotropy built in: as
in standard cosmology, the observed large-scale isotropy and
homogeneity remains to be explained.
\end{abstract}

\pacs{PACS numbers: 98.80.Cq}

]

\section{Introduction}
The brane world scenario has recently received attention as a
possible string inspired cosmology (see~\cite{roy2} for a review).
In this scenario the observable universe is a 4-dimensional (4-d)
slice, the brane, in a higher dimensional spacetime, the bulk. Here we consider 
the particular implementation  developed in~\cite{SMS}
in order to generalise a previous model by Randall and Sundrum
\cite{RS}, where the bulk is 5-dimensional (5-d) and contains only a
cosmological constant, assumed to be negative.  In this context
various authors~\cite{MS,SC2,SVF,SC1,coley2,coley1} have considered
an homogeneous and anisotropic brane, finding an intriguing
result: unlike general relativity, where in general the cosmological
singularity is anisotropic, the past attractor for homogeneous
anisotropic models in the brane is a simple Robertson-Walker
model.  In particular, it is  found in~\cite{coley2,coley1} that this is
also true for Bianchi IX models, as well as for some inhomogeneous models.

In general relativity the Belinski-Lifshitz-Kalatnikov conjecture
\cite{BLK} suggests that the Bianchi IX behaviour in the vicinity of
the singularity is general, i.e.\ that the approach to the
cosmological singularity in a generic inhomogeneous universe model
should locally be the same as in Bianchi IX.

Building on this and on the fact that Bianchi IX models in the brane have
isotropic singularities, it has been suggested in~\cite{coley2,coley1}  that
the isotropic singularity could be a generic feature of brane cosmological
models.

A well known problem of cosmology is to explain the very high degree of
isotropy observed in the CMB. In a theory such as general relativity, where
isotropy is a special rather than generic feature of cosmological models, we
need a dynamical mechanism able to produce isotropy.
Inflation was proposed, among other reasons, as a way to isotropise the
universe. Inflation is effective in this sense,
\footnote{There are perturbative proofs of the cosmic no-hair 
conjecture, i.e.\ classical perturbations in inflationary models 
with a scalar field or cosmological constant are swept out, as well 
as (local) proofs for homogeneous and inhomogeneous models (see e.g.~\cite{BMT} 
and references therein).} but  it needs homogeneous enough initial 
data in order to start at all~\cite{KT}. Although one can adopt the 
view that one such homogeneous enough patch in an
otherwise inhomogenous initial universe is all we need to explain what we
observe, this seems somehow unsatisfactory: the isotropy problem remains
open in standard cosmology.

If the conjecture in~\cite{coley2,coley1} could be proved correct, brane
cosmology would have the very attractive feature of having isotropy built
in. Inflation in this context would still be the most likely  way of
producing the fluctuations seen in the CMB, but there would be no need of
special initial conditions for it to start.\footnote{See \cite{BM} for CMB limits on the  anisotropy in brane cosmology.}  Also, Penrose conjecture~\cite{penrose} on
gravitational entropy and an initially
vanishing Weyl tensor would be satisfied, cf.~\cite{tod}.

Unfortunately for the brane scenario, we prove here that the past attractor of homogeneous models
fund in~\cite{coley2,coley1} is unstable in the past to generic (i.e.\
anisotropic and inhomogeneous) perturbations.  As in general relativity,
there is a decaying mode in scalar perturbations that grows unbounded in the
past and that signal, in the context of linear perturbation theory, that
anisotropy also grows unbounded as $t\rightarrow 0$.

In order to prove this, we specifically look at perturbation modes of the
dimensionless shear $\sigma/H$, and show that there is a decaying mode
(growing in the past) in this quantity.
Ours is a large-scale analysis, at a time when  physical scales of
perturbations are much larger than the Hubble radius,  $\lambda\gg H^{-1}$
(equivalent to neglecting Laplacian terms in the evolution equations). This may
seem restrictive, but this is not the case for the non-inflationary perfect
fluid models tha are relevant to our discussion. Indeed in this case any wavelength $\lambda$,
smaller than $H^{-1}$ at a given time, becomes much larger than
$H^{-1}$ at earlier enough times. Because of this crucial property 
of perturbations for non-inflationary models our analysis is completely 
general, i.e.\ valid for any $\lambda$ as $t\rightarrow 0$.

For the most part we follow the notation and convention of
\cite{roy2,roy1,GM}. In section 2 we briefly summarise those results on
general brane dynamics that are relevant for the following discussion
(see~\cite{roy2,SMS,roy1} for more details and other references).
In section 3 we present the equations for scalar perturbations and
derive the large-scale evolution for the gauge-invariant density
perturbation
variable in the high energy limit. In particular we highlight the
decaying  mode that grows unbounded in the past. In section 4 we show
that a corresponding mode in the  dimensionless shear
$\sigma/H$ also grows in the past. In section 5 we draw our
conclusions.

\section{Brane dynamics}
The implementation of the brane-world scenario considered in~\cite{SMS}
assumes that the whole spacetime is 5-d and governed by 5-d Einstein  field
equations:
\begin{equation}
\widetilde{G}_{AB}=\widetilde{\kappa}^2\left[-\widetilde{\Lambda}
\widetilde{g}_{AB} +\delta_{AB}(\chi)\left(-\lambda g_{AB}
+T_{AB}\right)\right].
\label{eq:5d}
 \end{equation}
These represent a
4-d brane at $\chi=0$ living in a bulk with metric
$\widetilde{g}_{AB}$ and cosmological constant $\widetilde{\Lambda}$;
$\widetilde{\kappa}^2$ is the 5-d gravitational constant,
$\lambda$ is the brane tension, ${g}_{AB}$ and $T_{AB}$ are
respectively the metric and the energy-momentum on the brane.  The
4-d field equations induced on the brane are derived
geometrically from (\ref{eq:5d}) assuming a $Z_2$ symmetry with the
brane as the fixed point, leading to modified Einstein equations with
new terms representing bulk effects:
\begin{equation}
G_{\mu\nu}=-\Lambda g_{\mu\nu}+\kappa^2
T_{\mu\nu}+\widetilde{\kappa}^4S_{\mu\nu} - {\cal E}_{\mu\nu}\,,
\label{2}
\end{equation}
where as usual  $\kappa^2=8\pi/M_{\rm p}^2$. The
various physical constants and parameters appearing in the equations
above are not independent, but related to each other by
\begin{equation}
\lambda=6{\kappa^2\over\widetilde\kappa^4} \,, ~~ \Lambda =
{\textstyle{1\over2}}\widetilde\kappa^2\left(\widetilde{\Lambda}+
{\textstyle{1\over6}}\widetilde\kappa^2\lambda^2\right)\,.
\label{3}
\end{equation}
The tensor $S_{\mu\nu}$ represents non-linear matter corrections
given by
\begin{eqnarray}
\lefteqn{ S_{\mu\nu}={\textstyle{1\over12}}T_\alpha{}^\alpha T_{\mu\nu}
-{\textstyle{1\over4}}T_{\mu\alpha}T^\alpha{}_\nu }  \nonumber \\
& &
~~~~~~~+ {\textstyle{1\over24}}g_{\mu\nu} \left[3 T_{\alpha\beta}
T^{\alpha\beta}-\left(T_\alpha{}^\alpha\right)^2 \right]\,.
\label{3'}
\end{eqnarray}
 ${\cal E}_{\mu\nu}$ is the projection on the brane of the
5-d Weyl tensor;  although the whole dynamics is
5-d and given by (\ref{eq:5d}), from the 4-d
point of view ${\cal E}_{\mu\nu}$ is a  non-local source term term
that carries bulk effects onto the brane.
If $u^\mu$ is the 4-velocity of matter and
$h_{\mu\nu}=g_{\mu\nu}+u_\mu u_\nu$ projects into the
comoving rest space, one can
decompose ${\cal E}_{\mu\nu}$ as~\cite{roy2,roy1}:
\begin{equation}
{\cal E}_{\mu\nu}={-6\over\kappa^2\lambda}\left[{\cal
U}\left(u_\mu u_\nu+{\textstyle {1\over3}} h_{\mu\nu}\right)+{\cal
P}_{\mu\nu}+{\cal Q}_{\mu}u_{\nu}+{\cal Q}_{\nu}u_{\mu}\right]\,,
\label{6}
\end{equation}
effectively as if it was a trace-less energy momentum
tensor with energy density ${\cal U}$, energy flux ${\cal Q}_{\mu}$
and anisotropic pressure ${\cal P}_{\mu\nu}$.
The brane energy-momentum tensor separately satisfies the
conservation equations, $\nabla^\nu T_{\mu\nu}=0 $. Assuming
a perfect fluid (or minimally coupled scalar field) we have the usual results:
\begin{eqnarray}
&&\dot{\rho}+\Theta(\rho+p)=0\,,\label{pc1}\\ && \D_\mu
p+(\rho+p)A_\mu =0\,,\label{pc2}
\end{eqnarray}
where a dot denotes $u^\nu\nabla_\nu$, $\Theta=\D^\mu u_\mu$ is the
volume expansion, $A_\mu=\dot{u}_\mu$ is the 4-acceleration, and
$\D_\mu$ denotes the spatially projected covariant derivative.  The
contracted Bianchi identities on the brane then imply that the
projected Weyl tensor ${\cal E}_{\mu\nu}$ and $S_{\mu\nu}$ obey the
constraints
\begin{equation}
\nabla^\mu{\cal E}_{\mu\nu}={6\kappa^2\over\lambda}\nabla^\mu
S_{\mu\nu}\,, \label{5}
\end{equation}
which  show how the non-local bulk effects are sourced by the
evolution and spatial inhomogeneity  of the brane matter content.
Finally, using Eqs.~ (\ref{2})-(\ref{pc2}), these can be turned in
propagation equations for the non-local energy density ${\cal U}$
and energy flux ${\cal Q}_\mu$. Neglecting  terms quadratic in
covariant  variables that are gauge-invariant  perturbations with respect to
a Robertson-Walker isotropic background\cite{roy1,EB,BDE}, these
are:\footnote{Strictly speaking, the variables defined
in~\cite{EB,BDE,DBE} and those defined in the same way in the
brane context~\cite{roy2,roy1,Leong} are 4-d; they are however easily
generalised to 5-d. Bardeen like variables~\cite{bardeen,KS} have been defined in 5-d in order to carry out a  brane-bulk analysis, e.g.\ see~\cite{BMW}, but the relation between these and the covariant ones used here has not yet been established; cf.~\cite{BDE,DBE} for this relation in 4-d general relativity. }
\begin{eqnarray}
&& \dot{\cal U}+{\textstyle{4\over3}}\Theta{\cal U}+\D^\mu{\cal
Q}_\mu  =0 \,, \label{lc1'} \\
&& \dot{\cal Q}_{\mu}+4H{\cal Q}_\mu
+{\textstyle{1\over3}}\D_\mu{\cal U}+{\textstyle{4\over3}}{\cal
U}A_\mu \nonumber \\ 
&& ~~~~~~~~~~~~~~~~
+\D^\nu{\cal P}_{\mu\nu}
=-{\textstyle{1\over6}} \kappa^4(\rho+p) \D_\mu \rho
\,,\label{lc2'}
\end{eqnarray}
where $H=\dot{a}/a$ ($={1\over3}\Theta$) is the Hubble expansion of
the background. Using a standard decompositions of perturbations,
the  anisotropic term ${\cal P}_{\mu\nu}$ carries in general
contributions from scalar, vector and tensor modes; the latter
however,  satisfying $\D^\nu{\cal P}_{\mu\nu}=0$, does not
contribute the non-local conservation equations above.

In the background, the Raychaudhuri equation is
\begin{eqnarray}
\lefteqn{\dot{H}=-H^2-{\kappa^2\over6}\left[\rho+3p+{\rho\over \lambda}
(2\rho+3p)\right] } \nonumber \\
&& ~~~~~~~~~~~~~~~~~~~~+{1\over3}\Lambda- {2\over
\kappa^2\lambda }{\cal U}_o\left({a_o\over
a}\right)^4\,,\label{bray}
\end{eqnarray}
where the solution for ${\cal U}$ follows from Eq.~(\ref{lc1'})
(considering the zero order background only), $a_o$ is the initial
scale factor and ${\cal U}_o={\cal U}(a_o)$. The first integral of
this equation is the generalized Friedmann equation on the brane
($K=0,\pm1$):
\begin{equation}\label{f}
H^2={\kappa^2\over3} \rho\left(1 +{\rho\over2\lambda}\right) +
{1\over3}\Lambda -{K\over a^2}+ {2\over \kappa^2\lambda } {\cal
U}_o\left({a_o\over a}\right)^4\,.
\end{equation}
 The high energy regime is defined by $\rho \gg \lambda$.
  In this limit   one obtains   flat models dominated by the non-linear
$\rho$ term~\cite{BDL},
\begin{equation}
H^2=\frac{\kappa^2}{6\lambda}\rho^2\,, ~~~~  a=
\left(\frac{t}{t_o}\right)^\frac{1}{3(1+w)}\,, \label{Fb}
\end{equation}
 where we fix an arbitrary initial
condition by choosing  $a_{\rm o}=a(t_o)=1$, and  as usual $w=p/\rho$. This
models are represented by  a stationary (equilibrum)  point, denoted ${\cal
F}_b$, in the phase space of homogeneous Bianchi models consider
in~\cite{coley2,coley1}, as well as in the phase space of a special class of
inhomogeneous $G_2$ cosmological models. In both cases ${\cal F}_b$ is found
to be the source, or past attractor, for the generic dynamics for $w> 0$
($w=0$ is also included in the homogeneous case), consistently
with~\cite{MS,SC2,SVF,SC1}.
 
Finally, we note that  the condition for
inflation in general is~\cite{MWBH}
\begin{equation}\label{inf}
w<-{1\over3}\left({2\rho+\lambda\over \rho+\lambda} \right)\,.
\end{equation}
 As $\rho/\lambda\to\infty$ this becomes $w<-{2\over3}$, while the
general relativity condition $w<-{1\over3}$ is recovered as
$\rho/\lambda\to 0$.

\section{Density perturbations}
Scalar gauge-invariant  perturbations can be described
using  covariantly defined variables (see~\cite{EB,BDE} and
references therein). In the brane scenario this formalism has been
developed in~\cite{roy2,roy1} (see also~\cite{Leong}); here we shall follow the
same approach, with minor modifications, and we refer to these papers for
definitions. 

 We can  completely characterize scalar perturbations on
the brane with  four variables,
$\Delta$, $C$, $U$ and $Q$, representing respectively the matter
density perturbation, a convenient 3-curvature perturbation, the
perturbation of the Weyl energy density, and a Weyl energy flux variable
related to ${\cal Q}_\mu$ in (\ref{6}) (all
3-quantities are defined with respect to a single 4-velocity field,
that of matter $u^{\mu}$). The dynamics of these quantities is
given by
\begin{eqnarray}
\lefteqn{
\dot{\Delta} =\left[3wH -{\kappa^2 \rho (1+w) \over 2 H}
\left(1+{\rho\over\lambda} \right)\right] \Delta } \nonumber \\
& & ~~~~~~  + {(1+w) \over 4a^2
H}C -{ \rho (1+w) \over 2 H} \left( {6\over \lambda\kappa^{2}}\right)
U  \,,\label{Ddot}\\
\lefteqn{
\dot{C}=\left({4 a^{2}H c_{\rm s}^2\over 1+w}\right)
\D^2\Delta } \nonumber \\
& & ~~~~~ - \left({12 a^{3} \rho\over\lambda\kappa^{2}}\right)
\D^{2}Q
-\left({72 a^{2}H c_{\rm s}^2\over 1+w}\right)
\left({{\cal U}\over\lambda\kappa^{2}}\right) \Delta\,,
\label{Cdot}\\
\lefteqn{
  \dot{U} =(3w-1)HU} \nonumber \\
& & ~~~~~ - \left({4c_{\rm
s}^2\over 1+w}\right)\left({{\cal U}\over\rho}\right) H\Delta
-\left({4{\cal U}\over3\rho}\right)
Z-a\D^2Q\,,\label{Udot}\\
\lefteqn{
 \dot{Q}
=(1-3w)HQ-{1\over3a}U-{\textstyle{2\over3}} a\D^2P }
\nonumber\\
& & ~~~~~
+{1\over6a}\left[ \left({8c_{\rm s}^2\over
1+w}\right){{\cal U}\over\rho}-\kappa^4
\rho(1+w)\right]\Delta\,.\label{Qdot}
\end{eqnarray}
Note that this system of equation is homogeneous in the four chosen
variables, except for the $P$ term in (\ref{Qdot}); this represent the
contribution from the anisotropic Weyl stress, and since there is no
evolution equation for $P$, in general one should either determine $P$
from the full \mbox{5-d} dynamics, or make some ansatz; otherwise, in general the
system above is not closed and one cannot find solutions.
Finally, the variable $Z$ in (\ref{Udot}),
used in
\cite{roy2,roy1,GM} and characterising the perturbation of the
expansion,
is related by $\Delta$, $U$ and $C$ by the constraint
\begin{equation}
C=2\kappa^{2}a^{2}\rho\left(1 + {\rho \over\lambda}\right)\Delta
+{12\over \lambda\kappa^{2}} a^{2}\rho U -4a^{2} H Z\;,
\label{Cconstr}
\end{equation}
 arising from the Gauss-Codazzi constraint in the brane.

In the following we want to study  the stability properties of the  models ${\cal F}_b$, Eq.~(\ref{Fb}), against generic inhomogeneous and anisotropic perturbations, for values of $w\geq 0$.
We see from (\ref{inf}) that these models   are non-inflationary for  $w\geq 0 $. Thus,  as we said in the introduction, for any $\lambda$ we only need to
study the large-scale evolution of the variables above.
We can either  neglect the Laplacian terms
in Eqs.~(\ref{Ddot})-(\ref{Qdot}), or use a harmonic expansion
(Fourier in our flat space case) and neglect terms
$H^{-2}/\lambda^{2} \ll 1$: the resulting equations are the
same.

 Fortunately, in the large-scale limit  the $\D^2 Q$ term in (\ref{Udot}) is negligible,
and one obtains a closed system for the density perturbations $\Delta$ and
$U$ and the curvature variable $C$. Besides, to the extent that  the $P$
contribution to (\ref{Qdot}) is also negligible in this limit, cf.\cite{LMSW}, $Q$ can also be determined~\cite{roy2,roy1,GM}. In
addition, we now restric our analysis to the
case ${\cal U}=0$: we will comment on the reliability  of this
assumption in the conclusions. 

Introducing the new variables
$\tilde{U}=U/(\kappa^{4}\rho)$ and $\tilde{Q}=aHQ/\kappa^4\rho$
and denoting with a prime the derivative with respect to
$d\tau=d\ln(a)$, we have
\begin{eqnarray}
 {\Delta}^\prime& =& \left[3w -{\kappa^2 \rho (1+w) \over 2 H^{2}}
\left(1+{\rho\over\lambda} \right)\right]\Delta  \nonumber \\
 & & + {(1+w) \over 4a^2
H^2}C -{ 3(1+w)\kappa^{2}\rho^{2} \over \lambda H^2}  \tilde{U}
\,,\label{Dp}\\
{C}^\prime&=&0\,,
\label{Cp}\\
 {\tilde{U}}^\prime &=&2(3w+1)\tilde{U}\,.\label{Up}
\\
 {\tilde{Q}}^\prime&=&\left(2-3w\right)\tilde{Q}-
{1\over 3}\tilde{U}-{1\over 6} (1+w)\Delta\;. \label{DQ}
\end{eqnarray}
We see therefore that the evolution in the long wavelength limit, with
${\cal U}=0$, considerably simplifies: it is described by a single
first
order equation for the density perturbation and two first integrals
arising from (\ref{Cp})-(\ref{Up}).
The first, $C_{\rm o}$, represent the large-scale constant spatial
curvature perturbation; the second, $\tilde{U}_{\rm
o}=\tilde{U} a^{-2(1+3w)}$, is a first integral for the Weyl energy
density perturbation.  These source the density perturbation;
therefore, $\Delta$ has three modes, as opposed to the two arising in
GR in the case of a single fluid.  The first two are analogous to the
GR ones: the first is the  mode arising from the $C_{\rm
o}=\tilde{U}_{\rm o}=0$ initial condition, typically a ``decaying
mode'', the second is the curvature adiabatic mode generated by
$C_{\rm 0}$; the third mode is a peculiarity of the brane scenario, is
generated by $\tilde{U}_0$, and represent an isocurvature (or entropy)
perturbation.  Although this is not particularly clear from the
treatment used here and in~\cite{roy2,roy1,GM}, this mode is indeed
due to the different intrinsic 4-velocities that the ``Weyl fluid''
and matter have in the perturbed spacetime (which is the cause of the
presence of the energy flux variable $Q$).  In complete analogy with
the case of two fluids in GR, this then generates a ``relative entropy
perturbation'' (see e.g.~\cite{DBE,KS} and~\cite{GM}).
Finally, and again as in GR, we remark that the adiabatic and
isocurvature perturbations evolve independently as described above
only in the long wavelength limit: it is indeed clear from Eqs.~(\ref{Ddot})-(\ref{Qdot}) than in general the two modes are coupled
and non-vanishing even starting from $C_{\rm o}=\tilde{U}_{\rm o}=0$.
Conversely the decaying mode should be seen as arising from special
initial conditions that lead to vanishing values of $C$ and
$\tilde{U}$ when $\lambda\gg H^{-1}$. Usually this mode is decaying
forward in time, and therefore neglected in structure formation
studies, while the growing mode is the interesting one. In studying
the question of homogeneity and isotropy of the brane at early times
we are interested in running the equations above backwords
in time, and it is the decaying mode that plays a crucial role  for most
values of $w$, as we are now going to show.

Let us now restrict our analysis to the high-energy regime that
dominates at very early times, when the background is given by the models ${\cal F}_b$,  Eq.~(\ref{Fb}).
Using $C_{\rm o}$ and $\tilde{U}_{\rm o}$, in the
 limit $\rho\gg \lambda$ the evolution of $\Delta$ is determined by
the
simple equation
\begin{equation}
{\Delta}^\prime=-3\Delta+\frac{9}{4}(1+w)^3C_{\rm o}
a^q-18(1+w)\tilde{U}_{\rm o}a^r\;,
\end{equation}
from which the three density perturbation modes are
immediately determined. They are
\begin{equation}
\Delta=\Delta_{\rm o}a^p+\frac{9}{4}\frac{(1+w)^3}{6w+7}C_{\rm
o}a^q-\frac{18(1+w)}{6w+5}\tilde{U}_{\rm o}a^r\;,
\label{Deltasol}
\end{equation}
where
\begin{eqnarray}
p=-3\;,&~~q=6w+4\;,&~~r=2(1+3w)\;, \\
&q<0 \Leftrightarrow
w<-\frac{2}{3}\;,~~~~ &r<0 \Leftrightarrow w<-\frac{1}{3}\;,
\end{eqnarray}
and $\Delta_{\rm o}$ is the constant of integration associated
with the decaying mode. This shows that independently of the
value of $w$, there is always a large-scale  mode that
grows unbounded in the past.

The solution for $\tilde{Q}$ can also be easily determined:
\begin{eqnarray}
\lefteqn{ 
\tilde{Q}=\tilde{Q}_{\rm
o}a^s+\frac{1}{6}\frac{1+w}{5-3w}\Delta_{\rm o}a^p } \nonumber \\
 & & ~~~ -\frac{3}{8}\frac{(1+w)^4}{(7+6w)(2+9w)}C_{\rm o}a^q
+\frac{1}{27}\frac{(2+3w)^2}{w(5+6w)}\tilde{U}_{\rm o}a^r
\label{Qsol}
\end{eqnarray}
for $w\neq 0$ and
\begin{equation}
\tilde{Q}=\tilde{Q}_{\rm o}a^{2}+\frac{1}{30}\Delta_{\rm o}a^{-3}
-\frac{3}{112}C_{\rm o}a^{4}+\frac{4}{45}\tilde{U}_{\rm
o}a^{2}\ln a
\end{equation}
for $w=0$, where
\begin{equation}
s=2-3w\;, ~~~~ s<0 \Leftrightarrow w>\frac{2}{3}\;,
\end{equation}
and $\tilde{Q}_{\rm o}$ is a constant of integration representing the
homogeneous solution to equation (\ref{DQ}).

\section{The Expansion Normalised Shear}
Like in the standard GR case, all the gauge-invariant geometric
and kinematic quantities can be expressed in terms of $\Delta$
\cite{Goode} in the context of linear perturbation theory. The key
covariant variable related to the issue of isotropization in the past is
the expansion normalised shear~\cite{GW},
\begin{equation}
\Sigma_{ab}=\frac{\sigma_{ab}}{H}\;.
\end{equation}
The scalar contribution to this quantity is obtained by taking its
total spatial divergence~\cite{BDE}:
\begin{equation}
\Sigma=\frac{a^2D^a D^b\sigma_{ab}}{H}\;.
\end{equation}
Using the shear constraint equation (equation (97) in~\cite{roy2})
it is easy to show that $\Sigma$ can be expressed in terms of
$\Delta$, $C$ $\tilde{U}$ and  $\tilde{Q}$:
\begin{equation}
\Sigma=2\Delta-\frac{1}{4a^2H^2}C+12\tilde{U}+36\tilde{Q}\;,
\end{equation}
and using (\ref{Deltasol}) and (\ref{Qsol}) together with the
solutions
for $\tilde{U}$ and $\tilde{Q}$ we obtain
\begin{eqnarray}
\lefteqn{ \Sigma=\frac{16}{5-3w}\Delta_{\rm
o}a^p-\frac{3}{4}\frac{(1+w)^2(99w^2+153w+40)}{(6w+7)(2+9w)}C_{\rm
o}a^q } \nonumber \\ && ~~~~~~~~~~~~~~~~~~~
+
\frac{8}{3}\frac{18w^2+15w+2}{w(62+5)}\tilde{U}_{\rm o}a^r +36\tilde{Q}_{\rm o}a^s
\label{shear1}
\end{eqnarray}
for $w\neq 0$ and
\begin{equation}
\Sigma=\frac{16}{5}\Delta_{\rm
o}a^{-3}-\frac{51}{28}C_{\rm o}a^{4} +\frac{24}{5}\tilde{U}_{\rm
o}\left(1+2\ln a\right)a^{2} +36\tilde{Q}_{\rm o}a^{2}
\label{shear2}
\end{equation}
for $w=0$.
The presence of the decaying mode $p=-3$ in (\ref{shear1}) and
(\ref{shear2}) proves that $\Sigma$ grows in the past. This completes our
proof that the past attractor ${\cal F}_b$ of homogeneous models is unstable in the past
against anisotropic and inhomogeneous perturbations.
\section{Conclusions}
From a dynamical system point of view the past attractor ${\cal F}_b$ for
brane homogeneous cosmological models found in~\cite{coley2,coley1} is a
fixed point in the phase space of these models. This phase space  may be
though of as an invariant submanifold  within an higher dimensional phase
space for more general inhomogeneous  models. The conjecture
in~\cite{coley2,coley1} is equivalent to saying, in this dynamical system
language, that ${\cal F}_b$ is the (local)  past attractor for generic
trajectories in this higher dimensional phase space. Our analyisis can be
seen as an exploration of the neighborhood of ${\cal F}_b$ out of the
invariant submanifold explored in~\cite{coley2,coley1}. We have found that
${\cal F}_b$ is unstable in the past to generic anisotropic and
inhomogeneous perturbations of non-inflationary perfect fluid models with
$p=w\rho$, for any value of $w$, using a large-scale $\lambda\gg H^{-1}$
approximation that we have motivated in the introduction and is not
restricting the validity of our analysis. The instability of ${\cal F}_b$ we
have found is fundamentally  due, among other modes that may be stable or
not depending on the value of $w$, to a decaying mode in the density
perturbation that blows up in the past, $\Delta \sim a^{-3}$, in a way
independent of $w$ and that, like in general relativity, is the signal of
the unbounded growth of the dimensionless shear $\sigma/H$ as $t\rightarrow
0$, as proved in section 4.

We have considered here only the case of vanishing background Weyl energy
density, ${\cal U}=0$. This assumption considerably simplifies the analysis,
but it is an easy guess that our results will remain true for ${\cal U}\not
=0$. Indeed when ${\cal U}\not =0$ ${\cal F}_b$ still remains the past
attractor of the isotropic models whose stability in the past we want to
examine, as is clear from the Friedmann equation (\ref{f}). In other words,
our analysis is restricted to the invariant submanifold ${\cal U}=0$ of the
larger phase space with ${\cal U}\not =0$, but   this
submanifold is asymptotically stable against ${\cal U}\not =0$ perturbations.

Finally, it has recently been suggested~\cite{LSR} that the quantity ${\cal
U}_0$ in (\ref{f}) is only asymptotically a constant, while ${\cal U}_0 \sim
a^4$ at high enough energies. As it clear from (\ref{f}),  even more in this
case ${\cal F}_b$ still remains the relevant past attractor of isotropic
models. We believe therefore that our analyisis should remain valid also in
this case. 
A more complete analysis including this issue and therefore that of ${\cal
U} \not =0$ will be the subject of a future investigation.
\\

\noindent
{\bf Acknowledgements}
 ~MB thanks Roy Maartens, Carlos Sopuerta, Filippo Vernizzi and David Wands for useful
discussions, and the
Department of Mathematics and Applied Mathematics of the
University of Cape Town and the Albert   Einstein Institute (Golm) for
hospitality while part of this work was carried out. PKSD thanks for NRF (South Africa) for financial support.


\begin{references}


\bibitem{roy2}
R. Maartens,
in {\it  Reference Frames \& Gravitomagnetism}, eds. J. Pascual-Sanchez
et al., 93, (World Scientific 2001), {\tt gr-qc/0101059.}

\bibitem{SMS}
T. Shiromizu, K. I. Maeda, and M. Sasaki,
{ Phys. Rev.} D {\bf 62}, 024012 (2000).

\bibitem{RS}
L. Randall, R. Sundrum,
{ Phys. Rev. Lett.} {\bf 83}, 4690 (1999).


\bibitem{MS}
R. Maartens, V. Sahni and T. D. Saini,
{ Phys. Rev.} D {\bf 63}, 063509 (2001).


\bibitem{SC2}
A. Campos and C. F. Sopuerta,
{ Phys. Rev.} D {\bf 63}, 104012 (2001).

\bibitem{SVF}
M. G. Santos, F. Vernizzi, P. G. Ferreira,
{ Phys. Rev.} D {\bf 64}, 063506 (2001)

\bibitem{SC1}
A. Campos and C. F. Sopuerta,
{ Phys. Rev.} D {\bf 64}, 104011 (2001).

\bibitem{coley2}
A. A. Coley,
{ Class. Quant. Grav.} {\bf 19} L45 (2002).

\bibitem{coley1}
A. A. Coley,
 Phys. Rev. D, (to be published). {\tt hep-th/0110049.}


\bibitem{BLK}
V. A. Belinski, I. M. Khalatnikov and E. M. Lifshitz,
{ Adv. in Phys.}, {\bf 19}, 525 (1970).

\bibitem{BMT}
M.\ Bruni, F.\ C.\ Mena \& R.\ Tavakol,
 { Class. Quantum Grav.}, {\bf 19}, L23 (2002).


\bibitem{KT}
E. W. Kolb and M. S. Turner, { The Early Universe}, (Addison - 
Wesley, 1990).


\bibitem{BM}
J. D. Barrow and R. Maartens,
{ Phys. Lett.} B {\bf 532}, 153 (2002).

\bibitem{penrose}
R.\ Penrose, in {\it General Relativity: An Einstein centenary survey}, p. 581 eds. S.\ W.\ Hawking and W.\ Israel (Cambridge University Press, 1979).



\bibitem{tod}
P. Tod,
{ Class. Quant. Grav.} {\bf 7}, L13 (1990).



\bibitem{roy1}
R. Maartens,
{ Phys. Rev.} D {\bf 62}, 084023 (2000).

\bibitem{GM}
C. Gordon and R. Maartens
{ Phys. Rev.} D {\bf 63}, 044022 (2001).


\bibitem{EB}
G. F. R. Ellis and M. Bruni,
{ Phys. Rev.} D {\bf 40}, 1804 (1989).


\bibitem{BDE}
M. Bruni, P. K. S. Dunsby and G. F. R. Ellis,
{ Ap. J.} {\bf 395}, 34 (1992).



\bibitem{DBE}
P. K. S. Dunsby, M. Bruni and G. F. R. Ellis,
{ Ap. J.} {\bf 395}, 54 (1992).


\bibitem{Leong}
B. Leong, P. K. S. Dunsby, A. Challinor and A. Lasenby,
{ Phys. Rev.} D {\bf 65},104012 (2002). 


\bibitem{bardeen}
J. M. Bardeen,
{ Phys. Rev.} D {\bf 22}, 1882 (1980).



\bibitem{KS}
H. Kodama and M. Sasaki,
{ Prog. Theor. Phys.} {\bf 78}, 1 (1984).



\bibitem{BMW}
H. Bridgman, K. A. Malik and D. Wands,
{ Phys. Rev.} D {\bf 65}, 043502 (2002).


\bibitem{BDL}
P. Bin\'etruy, C. Deffayet and D. Langlois,
{ Nucl. Phys.} {\bf B 565}, 269 (2000).



\bibitem{MWBH}
R. Maartens D. Wands B. A. Bassett and I. P. C. Heard
{ Phys. Rev.} D {\bf 62}, 041301 (2000).

\bibitem{LMSW}
D. Langlois, R. Maartens, M. Sasaki and D. Wands,
{ Phys. Rev.} D {\bf 63}, 084009 (2001).

\bibitem{Goode}
S. W. Goode, { Phys. Rev. D} {\bf 39}, 2882 (1989).



\bibitem{GW}
S. W. Goode and J. Wainwright,
{ Class. Quantum Grav.} {\bf 2}, 99 (1985).



\bibitem{LSR}
D. Langlois, L. Sorbo and M. Rodr\'iguez-Mart\'inez, {\tt hep-th/0206146.}

\end{references}
\end{document}